\documentstyle[12pt]{article}

\advance\textwidth1cm
\advance\textheight2.5cm
\advance\oddsidemargin-1.4cm
\advance\topmargin-1.2cm

\def\semi{\mathbin{\hbox{\hskip2pt\vrule height 5.7pt depth -0.1pt
                width .25pt \hskip-2pt$\times$}}} 

\newcommand{\be}{\begin{equation}}
\newcommand{\ee}{\end{equation}}
\newcommand{\bea}{\begin{eqnarray}}
\newcommand{\eea}{\end{eqnarray}}
\newcommand{\pa}{\partial}
\newcommand{\ie}{{\it i.e.}}
\newcommand{\eg}{{\it e.g.}}
\newcommand{\cf}{{\it cf}}
\newcommand{\r}{representation}
\newcommand{\ir}{irreducible representation}

\newcommand{\aoo}{algebra of observables}
\newcommand{\occ}{observable content of the constraints}
\newcommand{\q}{quantization}
\newcommand{\qt}{quantum theory}
\newcommand{\obs}{observable}
\renewcommand{\a}{algebra}
\newcommand{\fo}{fundamental observables}
\newcommand{\cas}{Casimir}
\newcommand{\casel}{Casimir element}
\newcommand{\casinv}{Casimir invariant}
\newcommand{\casop}{Casimir operator}
\renewcommand{\O}{${\cal O}$}
\newcommand{\OT}{$\tilde{\cal O}$}
\newcommand{\QO}{${\cal QO}$}
\newcommand{\QOT}{${\cal Q} \tilde{\cal O}$}

\newcommand{\CQ}{$\bar{\cal C}_0$}
\newcommand{\OC}{$\cal OC$}

\renewcommand{\S}{$\cal S$}
\renewcommand{\c}{constraint}

\newcommand{\R}{{\bf R}}
\newcommand{\GLN}{GL_0(N,{\bf R})}
\newcommand{\GLZ}{GL_0(2,{\bf R})}

\newcommand{\glz}{gl(2,{\bf R})}

\newcommand{\x}{{\bf x}}
\newcommand{\y}{{\bf y}}
\newcommand{\e}{{\bf e}}
\newcommand{\K}{{\sf K}}
\newcommand{\J}{{\sf J}}
\newcommand{\X}{{\sf X}}
\newcommand{\D}{{\sf D}}
\newcommand{\T}{{\sf T}}
\newcommand{\M}{{\sf M}}
\newcommand{\V}{{\sf V}}
\newcommand{\U}{{\sf U}}
\renewcommand{\P}{{\sf P}}
\renewcommand{\L}{{\sf L}}
\renewcommand{\H}{{\sf H}}
\newcommand{\CZ}{{\sf I}_2}
\newcommand{\CD}{{\sf I}_3}
\newcommand{\CF}{{\sf I}_5}
\newcommand{\LG}{{\cal L} G}
\newcommand{\LH}{{\cal L} H}
\renewcommand{\l}{{\sf \Lambda}}

\newcommand{\F}{{\cal F}}

\newcommand{\Ce}{{\sf C}}
\newcommand{\G}{{\cal G}}
\newcommand{\Pe}{{\cal P}}

\begin{document}

\thispagestyle{empty}

\begin{titlepage}

\vspace*{-2.5cm}
\begin{flushright} {Universit\"at Freiburg\\THEP 96/17\\hep-th/9701112}
\end{flushright}
\vspace{0.5cm}
\begin{center}
{\LARGE\bf Algebraic Constraint Quantization\\[3mm]
and the\\[7mm]
Pseudo--Rigid Body$^*$}\\[1.5cm]
{\large Michael Trunk}\\[4mm]
{Universit\"at Freiburg\\
Fakult\"at f\"ur Physik\\
Hermann--Herder--Str.\ 3\\
D--79104 Freiburg\\
Germany\\[5mm]
e-mail: trunk@physik.uni-freiburg.de}\\[2.5cm]
{\bf Abstract}\\[5mm]
\end{center}

The pseudo--rigid body represents an example of a constrained system with
a non-unimodular gauge group. This system is used as a testing ground for
the application
of an algebraic \c\ quantization scheme which focusses on \obs\ quantities,
translating the vanishing of the \c s into \r\ conditions on the algebra
of \obs s. The \c\ which is responsible for the non-unimodularity of the
gauge group is shown not to contribute to the \obs\ content of the \c s,
\ie, not to impose any restrictions on the construction of the quantum
theory of the system. The application of the algebraic \c\ quantization
scheme yields a unique quantization of the physical degrees of freedom,
which are shown to form a realization of the so-called $CM(N)$--model of
collective motions.

\end{titlepage}

\section{Introduction}

The pseudo--rigid body \cite{DE 90} represents an example of a first class
constrained system with a complicated, non--Abelian and non-unimodular gauge
group. In the present paper this system will be used a testing ground for the
application of an algebraic concept for the implementation of classical phase
space \c s into the \qt, formulated heuristically in Ref.\ \cite{Tr 96}. The
aim of this algebraic concept is to circumvent the technical and conceptual
problems which beset the currently used methods for the \q\ of constrained
systems, where one has to impose requirements upon the ``quantization'' of
unphysical quantities, like the \c s or gauge conditions, which are
subsequently used to project the physical states out of an extended
Hilbert space. In contrast, as the connection between the classical
and quantum descriptions of a physical system is closest on the algebraic
level, the central idea of the algebraic concept consists in translating
the ``vanishing'' of the \c s into conditions which are imposed upon \obs\
quantities. This is achieved by treating the intrinsically defined \occ\
(see below) as supplying \r\ conditions for the identification of the
physical \r\ of the \aoo.

The example of the pseudo-rigid body has been chosen because
Duval, Elhadad, Gotay, \'Sniatycki, and Tuynman \cite{Tu 90,DE 91} have
shown that, when quantizing a first class constrained system with a
non-unimodular gauge group $H$ using the Dirac quantization procedure, the
usual invariance condition for projecting the physical states out of an
extended Hilbert space
\be \hat{\J}_\xi \Psi_{phys} = 0  \ee
must be replaced by a condition of quasi--invariance
\be \hat{\J}_\xi \Psi_{phys} = -\frac{i}2 {\rm tr}({\rm ad}_\xi ) \cdot
\Psi_{phys}  \ee
($\xi \in \LH$, the Lie algebra of $H$; $\hat{\J}_\xi$ is the operator
corresponding to the generator $\J_\xi$ of the action of the one-parameter
subgroup of $H$ generated by $\xi$). This means that the naive application
to constrained systems of, \eg, the Dirac \q\ scheme, which {\it does} require
the \q\ of unobservable quantities, {\it viz}.\ the \c s, can lead to
``demonstrably erroneous conclusions" \cite{DE 90}.

Therefore, this system
presents a touchstone for the algebraic concept for the \q\ of constrained
systems. It will be shown that its application reproduces, in particular, the
results of Ref.\ \cite{DE 90}. In contrast to Ref.\ \cite{DE 90}, our method
can be applied without having to take special care of the non-unimodularity of
the gauge group: the \c\ which causes the non-unimodularity does not possess
any \obs\ content. Nevertheless, the discussion of this example gives rise to
a more precise and more widely applicable formulation of the concept.

The plan of the paper is as follows. In Sec.\ II the algebraic concept for
the implementation of classical phase space \c s into the \qt\ is
formulated. In Sec.\ III the pseudo-rigid
body is introduced in an arbitrary number $N$ of space dimensions. The
identification of the algebra of \obs s leads to the {\it CM(N)}--model of
collective motions. In Sec.\ IV the discussion is specialized to the dimensions
$N=2$ and $N=3$, the \obs\ content of the constraints is determined, and a
short description of the free dynamics is given. Finally, the quantization
of the system according to the algebraic concept, carried out in Sec.\ V,
yields a unique identification of the physical \r\ of the algebra of \obs s.

\section{Algebraic constraint quantization}

\subsection{Heuristic considerations}

To begin with, quantization is understood as the construction of the
quantum \aoo, starting from the classical \aoo, and the identification
of that irreducible $*$--\r\ of it, which provides the description of
the physical system in question. The classical respectively quantum \aoo\
is a (graded, involutive) Poisson respectively commutator algebra
which is generated polynomially by a set of {\it fundamental} \obs s.
The physical \r\ of the quantum \aoo, as a commutator \a, is distinguished
by additional algebraic structures, like characteristic relations (with
respect to the associative product) between its elements, the values of
its \casel s, or extremal properties. Likewise, the physical realization
of the classical \aoo, as a Poisson \a, is distinguished by algebraic
structures which correspond to those of the quantum \aoo. Strictly
speaking, the set of additional algebraic structures, required to
determine its physical realization, should be considered as part of
the definition of the \aoo; it should be chosen minimally, such that
there is just one faithful \r\ of the so defined \aoo.

Now the question arises, how the construction of the \qt\ can be
affected by the presence of \c s, and how the condition of the
``vanishing'' of the \c s, \ie\ the gauge invariance of the system,
can be implemented into the \qt\ \cite{Anm 1}.

The first place, where the \c s could gain influence on the construction
of the \qt\ is the characterization of the \aoo\ by the commutators of
its elements. This influence can be excluded by requiring that the set of
\fo, which generates the classical \aoo, consists of proper \obs s, \ie\
that the \fo\ are gauge {\it in}variant and that their linear span does
not contain any generators of pure gauge transformations. For, in that
case, assuming for the moment that the reduced phase space does exist,
each element of the set of \fo\ induces a non-vanishing function on the
reduced phase space and the abstract Poisson \a s, generated polynomially by
these two sets of functions, are isomorphic. Consequently, the two realizations
of this \a\ can only differ as regards the set of algebraic structures
which are needed to characterize the \aoo\ beyond the commutation relations,
and the differences must disappear upon the vanishing of the \c s. For this
to be the case, there must exist dependencies between those elements of the
\aoo, in terms of which these algebraic structures are formulated, and certain
gauge invariant combinations of the \c s (\eg\ functional dependencies,
which, upon the vanishing of the \c s, induce relations between the elements
of the \aoo). So, the only possibility for the \c s to work their way into
the \qt\ is the existence of such dependencies, which represent the remaining
gauge redundancy that has not been eliminated by the choice of the set of \fo.

In the present work it will be assumed that the \r s of the quantum \aoo\
can be characterized by the values of its \casinv s alone \cite{Anm 3}
(otherwise further algebraic structures must be treated in essentially
the same way as the identities for the \cas s are treated below).
This is usually the case in physically relevant systems, especially if
the \fo\ form a Lie \a.

On this assumption the \c s can only have an impact on the construction of
the \qt\ if they impose restrictions on the values of the \cas s of the \aoo.
That is, there must exist functional dependencies, in the classical theory,
which allow to identify certain gauge invariant combinations of the \c s with
\casel s of the \aoo, and the condition of the vanishing of the \c s induces
identities which have to be fulfilled by the \casel s of the \aoo. These
identities allow to translate the gauge invariance of the classical system
into conditions on \obs\ quantities, which will be referred to as the {\it
\occ} \cite{Anm 4}. By imposing correspondence requirements, the operator
versions of the said gauge invariant combinations of the \c s can be
identified with central elements of the quantum \aoo, which permit to
formulate the \occ, and thus to implement the \c s, on the level of \qt.

\subsection{Formulation of the concept}

In the following the most important steps for the realization of the
algebraic concept for the implementation of classical phase space \c s into
the \qt\ will be enumerated.
This enumeration should not be misunderstood as a \q\ program that can
be applied algorithmically. Rather, it is meant as a statement of the
principles which, in one form or the other, should apply to the \q\ of
an arbitrary constrained system (with the restriction stated above), but
which has to be adapted in a case by case analysis to the concrete situation.

In any case the starting point is a gauge invariant Hamiltonian system
with phase space $P$ and a set of first--class \c s $\Ce_i$. The following
notation will be employed:\\
$\F = C^\infty(P)$ is the set of all smooth functions on $P$; \\
${\cal C}\: = \{ f\in\F | \exists g^i \in \F: f = \sum_i g^i {\sf C}_i \}$
is the set of all weakly vanishing functions on

$P$ (under suitable regularity assumptions on the \c s ${\sf C}_i$,
cf.\ \cite{HT 92}); \\
$\Pe = \{ f\in\F | \{ f ,{\sf C}_i \} = 0 \:\; \forall i \}$ is the strong
Poisson commutant of the \c s; \\
$\G = {\cal C}\cap\Pe$ is the set of gauge invariant combinations of the
\c s; the elements

of $\G$ will be referred to as the {\it generalized \casel s of the \c s}.

\begin{itemize}

\item  {\sf Observables}: Choose a set \OT\ $\subset \Pe\setminus\G$ of \fo,
such that $L(\tilde{\cal O}) \cap\G = \{0\}$ (where $L(\tilde{\cal O})$ is the
linear span of the elements of \OT). \OT\ must generate $\Pe$ weakly, \ie\
the (closure, with respect to a suitable norm, of the) polynomial \a\ over
\OT\ must coincide with $\Pe$ up to equivalence ($\Pe\ni f \approx g \in\Pe
\Longleftrightarrow f-g \in\G$), at least locally, cf.\ \cite{Is 84}. If
there is a $*$--involution on $\cal P$, \OT\ has to be closed with respect
to it.

Equipped with the Poisson bracket, $L(\tilde{\cal O})$ should possibly form
a Lie \a\ (which, in that case, will also be denoted by \OT). Otherwise the
Poisson brackets of the elements of \OT\ must be polynomial in the \fo. The
classical algebra of \obs s \O\ is the Poisson \a\ generated polynomially by
\OT\ (or, more generally, its completion with respect to a suitable norm).

The generators of the symmetry \a\ \S\ of the system should be contained
in \OT\ \cite{Anm 5}, and the Hamiltonian must be a simple polynomial
function of the elements of \OT.

\item  {\sf The Observable Content of the Constraints}: Determine the \occ,
\ie\ the functional dependencies between the \casel s of \O\ and the \c s,
and the conditions which are imposed upon the \casel s of \O\ by the vanishing
of the \c s.

The set of generalized \casel s of the \c s, respectively of the corresponding
\casel s of \O, which enter into the functional dependencies, will be denoted
by \OC.

\item  {\sf Identities}: In addition, one has to determine the identities for
the \casel s of \O\ which do not involve the \c s.

\item  {\sf Quantum Algebra of Observables}: Starting from the classical
Poisson \a\ \O\ one has to construct the commutator \a\ \QO, which
represents the analogue of \O\ on the level of \qt. The quantum \aoo\
\QO\ is generated polynomially by a set \QOT\ of fundamental \obs s.
The elements of \QOT\ are in one-to-one correspondence with the
elements of \OT. The algebraic structure of \QO\ is defined by the
commutators between its elements, which can be obtained derivatively
from the commutators between the \fo. The latter have to be inferred
from the Poisson brackets between the classical \fo\ by imposing
correspondence and consistency requirements.

Thus, the \obs\ linear (\ie, Lie) or linearizable symmetries of the
system should be preserved upon quantization, \ie\ this part of the
symmetry \a\ of the quantum system should be isomorphic to
that of the classical system (where the Poisson bracket has to be
replaced by ($-i/\hbar$) times the commutator). For the commutators
of arbitrary elements of \QOT\ this strict correspondence of commutators
and Poisson brackets cannot be required {\it a priori}. Rather,
($-i/\hbar$) times the commutators can differ from the Poisson
brackets by quantum corrections which are compatible with the
correspondence principle. The correction terms must be formed from
elements of \QO\ which possess a well-defined non-vanishing classical
limit, multiplied by explicit positive integer powers of $\hbar$.
Together with these explicit powers of $\hbar$ they must carry the
correct physical dimensions. The covariant transformation properties
(with respect to the linear or linearizable part of the symmetry \a)
of the \fo\ as well as of their commutators should be
preserved. If \OT\ carries a gradation or $*$--involution, these
structures must also be implemented into \QOT, and the commutator
structure must be compatible with them. Of course, if \OT\ is a
Lie \a, this should also be the case for \QOT. Then, \QO\ is the
enveloping \a\ of the Lie \a\ \QOT.

This deformation process may not result in the occurrence of additional
\obs s on the level of \qt\ which do not possess a classical analogue.

\item  {\sf Correspondence of Observables}: The expressions for specific
quantum \obs s, which correspond to given classical \obs s, and their
commutation relations have to be determined along the same lines.

\item  {\sf The Observable Content of the Constraints}:
The crucial step is the identification of those \casel s of \QO\ which
correspond to the elements of \OC\ (the principles for their identification
are the same as above), and of the conditions which express the \occ\ on the
level of \qt.

\item  {\sf Identities}:
In the same way one has to determine the form of those identities for
the \casel s of \QO\ which correspond to the classical identities for
the \casel s of \O\ which do not involve the \c s.

\item  {\sf Identification of the Physical Representation of \QO}:
Having established the algebraic structure of \QO, the physical \r\ is that
irreducible $*$--\r\ of \QO, in which the conditions, which express the \occ,
and the identities for the remaining \casel s of \QO\ are satisfied.

\end{itemize}

Note that we do not have to introduce an extended Hilbert space
(where the term ``extended Hilbert space" refers to any Hilbert space
containing unphysical states). Of course, the use of an extended Hilbert
space may facilitate the construction of \r s of \QO. But then the
Hilbert space will not be irreducible with respect to \QO, and the
selection of the physical subspace, \ie\ of the physical \r\ of \QO,
can be carried out with the help of conditions which are imposed on
\obs\ quantities.

\renewcommand{\a}{{\bf a}}

\section{The pseudo-rigid body}

In Ref. \cite{DE 90} the pseudo-rigid body (PRB) is defined kinematically
by specifying its
configuration space. Consider a distribution of mass points in $\R^N$, or a
continuous mass distribution, such that the volume (of the convex hull) is
non-zero. Let this object undergo collective linear deformations, \ie\ let
each mass point be subject to the same linear transformation. Then, starting
from an initial configuration, each other configuration can be obtained by
specifying the change in the position of the center of mass, \ie\ an
element of $\R^N$, and in the orientation, shape, and size of the body,
\ie\ an element of $\GLN$, the identity component of $GL(N,\R)$. That is,
the configuration space of the PRB is the group
\[G=\GLN \semi \R^N \]
($\semi$ denotes the semi-direct product).

Now suppose we are unable to detect different orientations,
sizes, and positions of the body, \ie\ the physical degrees of freedom
are the different shapes of the body. Then, the redundant degrees of freedom,
namely dilations, rotations, and translations, are described by the action
of the gauge group
\[ H=\left( \R^+ \times SO(N) \right) \semi \R^N \]
acting on $G$ from the left. The group $H$ is non-unimodular, the
non-unimodularity being effected by the action of the dilations ($\R^+$)
on the translations via the semi-direct product structure.

\subsection{Symplectic structure}

The configuration space $Q=G$ being an open subset of the space
$M(N,\R) \times \R^N$ ($M(N,\R)$ is the space of real
($N\times N$)--matrices), we can introduce global coordinates on $Q$, namely
the matrix elements $x_{ij}$, $1\le i,j \le N$, of the element $g=(x_{ij})
\in \GLN$, and the Cartesian coordinates $x_i$ of the vector $\x \in \R^N$.

The phase space $P=T^* G$ of the system can be identified with the product
$G \times \LG^*$ ($\LG^*$ is the dual of the Lie algebra $\LG$ of $G$) by
the trivialization
\[ T^*: G \times \LG^* \longrightarrow T^*G,    \qquad\qquad
   ((g,\x),(\alpha,\beta)) \longmapsto  \mu_{(\alpha,\beta)}(g,\x) \]
($(\alpha,\beta) \in \LG^* \simeq M(N,\R) \times \R^N$).
In the above coordinates the one-form $\mu_{(\alpha,\beta)} \in T^*G$ is
given explicitly by
\be \mu_{(\alpha,\beta)}(g,\x)=\alpha_{ij} dx_{ji} + \beta_i dx_i
    \label{triv} \ee
(repeated indices are summed over).

Denoting the coordinates on $\LG^*$ by $(p_{ij},p_i)$, the Liouville form
can be written as
\be \theta = p_{ij} dx_{ji} + p_i dx_i .\ee
The symplectic form is $\omega = - d\theta$.

\subsection{Infinitesimal generators}

For the purpose of later reference we will supply the expressions for
the infinitesimal generators of the action of the group $G$ on its cotangent
bundle $T^*G$ by left and right translations. Let
\[ \Phi^L : G \times G \longrightarrow G, \qquad\qquad
   ((h,\y),(g,\x)) \longmapsto (hg,\y + h \x) \]
be the left action of the group $G$ on itself by left translations, and let
$\Phi^{L\ast}: G \times T^*G \longrightarrow T^*G$ denote the canonical lift
of this action to $P=T^*G$. Then, in the above coordinates, the infinitesimal
generator for the action of the one-parameter subgroup generated by the
element $(A,\a)\in {\cal L} G \simeq M(N,\R) \times \R^N$ on $P$ is given
by
\be \J^L_{(A,\a)}(x,p) = a_{ij} x_{jk} p_{ki} + a_i p_i + p_i a_{ij} x_j.
\label{infgenL} \ee
Similarly, let
\[ \Phi^R:G \times G \longrightarrow G, \qquad\qquad
   ((h,\y),(g,\x)) \longmapsto (gh^{-1},\x- gh^{-1}\y) \]
be the left action of $G$ on itself by right translations, and let
$\Phi^{R\ast}$ denote the canonical lift to $P$. In this case the
infinitesimal generator is
\be \J^R_{(A,\a)}(x,p) = -x_{ij} a_{jk} p_{ki} - p_i x_{ij} a_j. \ee
The Lie algebras which are formed by the elements $\J^{L/R}_{(A,\a)}$,
$(A,\a)\in \LG$, are isomorphic to $\LG$
\be \big\{ \J^{L/R}_{(A,\a)},\J^{L/R}_{(B,{\bf b})} \big\} = \J^{L/R}_{[(A,\a),
    (B,{\bf b})]} = \J^{L/R}_{([A,B],A{\bf b} - B\a)}. \ee
As left and right translations commute, the corresponding generators satisfy
\be \big\{ \J^L_{(A,\a)},\J^R_{(B,{\bf b})} \big\} =0. \ee

\subsection{Constraints}

Let $D=E_N$ ($E_N$ is the ($N\times N$)--unit matrix), let $\{K_{ij}=-K_{ji}\}$
be the standard basis of $so(N)$, and $\{\e_i\}$ the standard basis of
$\R^N$. Then, the elements $(D,0)$, $(K_{ij},0)$, and $(0,\e_i)$ constitute
a basis of $\LH$, the Lie algebra of the gauge group $H$, and the infinitesimal
generators
\be \D:=\J^L_{(D,0)}, \qquad \K_{ij}:=\J^L_{(K_{ij},0)}, \qquad
    \P_i:=\J^L_{(0,\e_i)} \ee
span the constraint algebra ${\cal C}_0 \simeq \LH$.

\subsection{Fundamental observables}

One class of observables, which can readily be obtained, is given by the
generators of right translations $\J^R_{(A,\a)}$. As the generators $\J^R_
{(D,0)} = - \J^L_{(D,0)}$ and $\J^R_{(0,\e_i)} = - \sum_j x_{ji} \J^L_{(0,
\e_j)}$ are contained in the intersection $\Pe\cap{\cal C}$ of the strong
Poisson commutant of the \c s with the set of weakly vanishing functions,
we have to restrict ourselves to the $sl(N,\R)$ subalgebra of $\LG$, so
that only the observables $\J^R_{(A,0)}$, $A\in sl(N,\R)$, can be taken
to form part of \OT.

The action of $SL(N,\R)$ has to be supplemented by the ``translations along the
fibers" generated by an appropriate class of functions on the configuration
space $Q=G$ (cf. \cite{Is 84}). These functions can be chosen as
\be \X_{ij}=\X_{ji}=\lambda \big(\det g \big)^{-2/N} \big( g^t g \big)_{ij} =
    \lambda \big( \det g \big)^{-2/N} x_{ki} x_{kj}, \qquad \lambda > 0
\ee
($g^t$ is the transpose of $g\in \GLN$, the physical significance of the
parameter $\lambda$ will be determined in the next section, cf.\ the
discussion below eq.\ (\ref{ham2})). The functions $\X_{ij}$ are obviously
invariant under translations ($\R^N$) and under the action of $SO(N)$ from
the left. As they are homogeneous of degree zero in $x_{ij}$, they are also
invariant under dilations.

Being the elements of a symmetric matrix, the functions $\X_{ij}$
generate an action of the Abelian group $S(N)$ of real symmetric
($N\times N$)--matrices on $P=T^*G$. Choosing the matrices
\[ S_{ij}=S_{ji}=\frac12 \big( E_{ij} + E_{ji} \big), \qquad\qquad
   (E_{ij})_{kl}=\delta_{ik}\delta_{jl} \]
as a basis of $S(N)$, the action on $P$ of an element $S=s_{ij} S_{ij} \in
S(N)$ ($s_{ij}=s_{ji}$) is given by
\be \mu \longmapsto \mu - s_{ij} d\X_{ij} \ee
($\mu \in P)$, in coordinates
\be x_{kl} \longmapsto x_{kl} ,\qquad\quad
    p_{kl} \longmapsto p_{kl} - s_{ij} \frac{\pa \X_{ij}}{\pa x_{lk}}.\ee
The Poisson brackets of the functions $\J^R_{(A,0)}$ and ${\sf S}_{(S)}
= s_{ij}\X_{ij}$ close to form a realization of the Lie algebra $sl(n,\R)\semi
S(N)$. Denoting the elements of $sl(N,\R) \semi S(N)$ by $(A,S)$, the
infinitesimal generators by $\tilde{\J}_{(A,S)}:=\J^R_{(A,0)}+{\sf S}_{(S)}$,
the commutation relations read
\be \big\{ \tilde{\J}_{(A,S)},\tilde{\J}_{(B,T)} \big\} = \tilde{\J}_{[(A,S),
    (B,T)]} = \tilde{\J}_{([A,B],AT + TA^t - BS - SB^t)}. \ee
The corresponding group is $SL(N,\R) \semi S(N)$, with the group multiplication
law
\[ (g,S)(h,T)=(gh,S+gTg^t) .\]
This group is also denoted by $CM(N)$, the group of collective motions in
$N$ dimensions, and plays a prominent r\^ole in the description of collective
modes of multi-particle systems, \eg\ in nuclear physics
(cf. \cite{RR 76,WC 76}).

The action of $SL(N,\R)\semi S(N)$ on $P$ induces a transitive action on
the space of physical states of the system, \ie\ on the reduced phase space
$\bar{P} = T^*(H\setminus G)$. Consequently, the functions $\tilde{\J}_{
(A,S)}$ generate the algebra of all \obs\ quantities. Therefore, we shall
choose the Lie algebra $cm(N) = sl(N,\R) \semi S(N)$ as the Lie algebra
\OT\ of fundamental \obs s. In the next section further justification will
be given to this choice.

\section{The cases $N=2$ and $N=3$}

Having established the kinematical properties of the PRB and the Lie structure
of the algebra of fundamental \obs s, we now have to determine the Casimirs of
\O\ and the \obs\ content of the \c s. This will be done explicitly for the
space dimensions $N=2$ and $N=3$. For the sake of clarity and simplicity, the
case $N=2$ will be treated in detail, the largely analogous discussion of the
case $N=3$ can then be kept short.

\subsection{$N=2$}

In two space dimensions the configuration space is the group $G=\GLZ \semi
\R^2$, the phase space is $P=T^*G$, and the gauge group $H=\big( \R^+
\times SO(2) \big) \semi \R^2$.
The constraint algebra ${\cal C}_0$ is spanned by the functions
\be \D=x_{ij}p_{ji} + x_i p_i ,\qquad \K=(x_{1i}p_{i2} - x_{2i}p_{i1}) +
    (x_1 p_2 - x_2 p_1) ,\qquad \P_i=p_i  \ee
($i,j \in \{1,2\}$). Choosing the matrices
\be L_1=\frac12 \left( \begin{array}{rr} 1&0\\0&-1 \end{array} \right),\quad
    L_2=\frac12 \left( \begin{array}{rr} 0&1\\1&0  \end{array} \right),\quad
    L_3=\frac12 \left( \begin{array}{rr} 0&-1\\1&0 \end{array} \right)
    \label{lmu}\ee
as a basis of the Lie algebra $sl(2,\R)$, the Lie algebra \OT\ = $cm(2)=
sl(2,\R) \semi S(2)$ is generated by the functions
\bea \L_1 &=& -\frac12 \big( x_{i1}p_{1i} - x_{i2}p_{2i} \big)\\
     \L_2 &=& -\frac12 \big( x_{i1}p_{2i} + x_{i2}p_{1i} \big)\\
     \L_3 &=& -\frac12 \big(-x_{i1}p_{2i} + x_{i2}p_{1i} \big)\\
     \X_{ij} &=& \frac\lambda{{\sf d}} x_{ki} x_{kj} ,\qquad {\sf d}=\det g =
     \det (x_{ij}).
\eea
The Lie algebra $cm(2)$ is isomorphic to the Lie algebra $iso(2,1) = so(2,1)
\semi \R^3$ of the Poincar\'e group in ($2+1$) dimensions, as can be seen by
defining the functions
\be \M_{\mu\nu}:=\varepsilon_{\mu\nu\rho} \L_\rho, \quad \X_1:=\X_{12} ,\quad
    \X_2:=\frac12 (\X_{11}-\X_{22}) ,\quad \X_3:=\frac12 (\X_{11}+\X_{22}) \ee
($\mu,\nu,\rho \in \{1,2,3\}$) with the commutation relations
\be \big\{ \M_{\mu\nu},\M_{\mu\rho} \big\} = g_{\mu\mu} \M_{\nu\rho} ,\qquad
    \big\{ \M_{\mu\nu},\X_\mu \big\} = g_{\mu\mu} \X_\nu ,\qquad
    \big\{ \X_\mu,\X_\nu \big\} = 0 \ee
($g_{\mu\nu}$ = diag$(++-)$, the missing commutators vanish or can be obtained
from the listed ones using the antisymmetry of $\M_{\mu\nu}$). As the algebra
$iso(2,1)$ is more familiar than the algebra $cm(2)$, we will continue to work
with the former. This algebra possesses two Casimir invariants
\be \X^2=g^{\mu\nu} \X_\mu \X_\nu ,\qquad\qquad \CZ = \varepsilon^{\mu\nu
    \rho} \M_{\mu\nu} \X_\rho \label{cas2}\ee    
(greek indices are raised and lowered with $g^{\mu\nu}=g_{\mu\nu}$,
$\varepsilon_{123} = 1$).
For $\X^2 \le 0$ the sign of $\X_3$ is a third invariant. The \c\ algebra
${\cal C}_0$ does not possess any Casimirs for $N=2$.

Next we have to determine the identities which express the functional
dependencies between the Casimirs of \O\ and the \c s, and the identities
which are fulfilled by the Casimirs of \O\ without involving the \c s. We
find
\be \lambda \bar{\K} \equiv \CZ ,\qquad\quad \bar{\K} := \K - x_1 \P_2
    + x_2 \P_1  \label{crel2}\ee
and
\be \X^2 \equiv - \lambda^2 ,\qquad\qquad \mbox{sign}(\X_3)=1. \label{cid2}
\ee
The function $\bar{\K}$, which represents the \obs\ content of the \c s,
is not a Casimir of ${\cal C}_0$ but a generalized \casel\ of the \c s. This
property is not an effect of the non-unimodularity of the gauge group $H$,
but simply of its semi-direct product structure (which, in turn, causes also
the non-unimodularity).

It should be noted that the \c\ algebra ${\cal C}_0$, which is a semi-direct
product, can be replaced by an equivalent \c\ algebra $\bar{{\cal C}}_0 =
\big( \R^+\times so(2) \big) \times \R^2$, which is a direct product and is
generated by $\bar{\K}$, $\bar{\D}=\D-x_i \P_i$, and $\bar{\P}_i = \P_i$
(for $N=2$, $\bar{{\cal C}}_0$ is even Abelian). $\bar{\K}$ is a Casimir of
\CQ, that is it fulfills the definition of an element of \OC\ as it was given
in Ref.\ \cite{Tr 96}.

It should also be noted that the dilation \c\ $\D$ does not contribute
to the \obs\ content of the \c s, \ie\ it cannot have any impact on the
selection of the physical \r\ of the algebra of \obs s. Therefore it
would be quite unphysical to make the quantization of the system depend on
requirements which are imposed on the quantization of this quantity which,
from the point of view of the constrained system, is un\obs. It is one of
the merits
of Ref.\ \cite{DE 90} to have shown that a naive application of the Dirac
quantization scheme, which requires the quantization of this quantity, can
lead to substantially wrong results: it is the requirement that the operator
corresponding to $\D$ upon quantization be formally self-adjoint with respect
to an inner product on an extended Hilbert space containing unphysical states,
which necessitates the introduction of the correction term (2).

Of course, from the discussion of this single example nothing can be said
about the general case of a non-unimodular gauge group.

\subsection{Dynamics}

In this paragraph it will be shown how the Hamiltonian for the free dynamics
of the constrained system can be expressed as a function of the
$cm(2)$ \obs s, and how these \obs s can be given a physical interpretation.

Let the body be composed of $n \ge 3$ individual mass points with the same 
mass $m$, not all of them lying on the same line. Let the positions of the 
mass points, relative to the center of mass, at the time $t=0$ be $\x_a^0$,
$a= 1,2,\dots ,n$. Then the positions at time $t>0$, implementing the condition
that all mass points be subject to the same linear transformation, can be
written as
\be \x_a(t)=g(t) \x_a^0 ,\quad\qquad g(t) \in \GLZ ,\quad g(0)=E_2 .\ee
Define the mass--quadrupole tensor $q(t)$ by
\be q_{ij}(t) = \sum_{a=1}^n m x_{ai}(t) x_{aj}(t)
              = \big( g(t) q(0) g^t(t) \big)_{ij} \ee
and choose the basis in the center of mass frame such that $q^0:=q(0)$
becomes diagonal: $q^0$ = diag($q_1,q_2$). The kinetic energy of the
unconstrained motion of the body, relative to the center of mass, can be
expressed by $q^0$ and the time derivative $\dot{g}(t)$ of $g(t)$
\be \T=\frac12 {\rm tr} \big( \dot{g}(t) q^0 \dot{g}^t(t) \big) . \ee
Now, using a trivialization analogous to (\ref{triv}) for the tangent
bundle
\[ T: G \times {\cal L} G \longrightarrow TG,    \qquad
   ((g,\x),(A,\a)) \longmapsto v_{(A,\a)}(g,\x) =
   a_{ij} \frac{\pa}{\pa x_{ij}} + a_i \frac{\pa}{\pa x_i} \]
($A=(a_{ij}) \in gl(N,\R) \simeq M(N,\R)$, $\a \in \R^N$) and
neglecting the translations, we can
identify the tangent vector $\dot{g}(t)$ with an element $\Omega(t)$ of
the Lie algebra $\glz$, which has the same components as $\dot{g}(t)$.
Expanding $\Omega$ into the basis $D=E_2$ and $I_\mu := L_\mu$ (eq.\
\ref{lmu}) of $\glz$
\be \Omega = \frac12 \omega_D D + \omega_\mu I_\mu \ee
the kinetic energy becomes a function of the generalized ``angular" velocities
$\omega_D$ and $\omega_\mu$: $\T=\T(\omega_D,\omega_\mu)$. The generalized
``angular" momenta conjugate to $\omega_D$ and $\omega_\mu$
\be \J_D=\frac{\pa \T}{\pa \omega_D},\quad\qquad \J_\mu =
    \frac{\pa \T}{\pa \omega_\mu} \ee
are the infinitesimal generators of the group action by left translations,
\ie\ in the above coordinates on the group
\be \J_D = \frac12 \bar{\J}^L_{(D,0)} ,\quad\qquad \J_\mu =
    \bar{\J}^L_{(I_\mu,0)} \ee
where $\bar{\J}^L_{(A,0)} = a_{ij} x_{jk} p_{ki}$ (this is the same as
(\ref{infgenL}), neglecting the translations). The canonical Hamiltonian
for the free motion of the unconstrained system is a function of the momenta
$\J_D$ and $\J_\mu$
\be \H_0 = \omega_D \J_D + \omega_\mu \J_\mu - \T =
    \frac{(\J_D+\J_1)^2 + (\J_2+\J_3)^2}{2q_1} +
    \frac{(\J_D-\J_1)^2 + (\J_2-\J_3)^2}{2q_2}. \ee
Now we have to implement the constraints such that different points on the
same gauge orbit are dynamically identified. In accordance with Dirac
\cite{Di 64} this can be done by adding to $\H_0$ a combination of the
\c s with arbitrary functions as coefficients. Again neglecting the
translations, we obtain the extended Hamiltonian
\be \H_E = \H_0 + \lambda_D \bar{\D} + \lambda_K \bar{\K} \ee
($\bar{\D}$ and $\bar{\K}$ as above, note that $\bar{\D}=2\J_D$,
$\bar{\K}= 2\J_3$). Making use of the arbitrariness of the
multipliers $\lambda_D$ and $\lambda_K$, $\H_E$ can be brought into the form
\be \H_E = \H + \kappa_D \bar{\D}+\kappa_K \bar{\K} ,\quad\qquad \H=
    \frac1{2\theta} \J_\mu \J^\mu ,\qquad \theta := \frac{q_1 q_2}{q_1+q_2}\ee
where $\kappa_D$ and $\kappa_K$ are still arbitrary, and $\J^2=\J_\mu \J^\mu$
is the quadratic Casimir of the $sl(2,\R)$ subalgebra of $\glz$. Finally,
observing that we have the identity $\J^2 = \L^2$, where $\L^2 = \L_\mu \L^\mu$
is the quadratic Casimir of the $sl(2,\R) \simeq so(2,1)$ subalgebra of the
$cm(2) \simeq iso(2,1)$ algebra, the Hamiltonian $\H$ can be expressed as a
function of observables
\be \H=\frac1{2\theta} \L^2 \label{ham2}.\ee
The symmetry algebra of $\H$, \ie\ of the free motion of the constrained
system, is just the $sl(2,\R) \simeq so(2,1)$ subalgebra of $cm(2)$. For
other forms of collective Hamiltonians (especially for $cm(3)$) cf.\ the
cited literature \cite{RR 76,WC 76} and Refs.\ therein.

The fact that the symmetry algebra of the free motion is already
included in the algebra $cm(2)$ and that the Hamiltonian is a polynomial
function of its generators provides a dynamical justification for the
choice of the algebra $cm(2)$ as the Lie algebra \OT\ of fundamental \obs s.
Furthermore, the functions
$\X_{ij}$ can also be given a physical interpretation. They can be thought
of as (redundant) coordinates on the orbit through the point $\lambda E_2
\in S(2)$ under the right action $S \longmapsto g^t S g$ of $SL(2,\R)$ on
$S(2)$. This orbit consists of all positive definite symmetric $(2 \times
2)$--matrices with determinant $\lambda^2$, and, with $\lambda =\sqrt{\det
q^0}$, can be identified with the space of all mass--quadrupole tensors of
determinant $q_1 q_2$. Moreover, as the quantity $\sqrt{q_1 q_2}$ is equal
to the product of the mass and the volume of the body, the invariant $\X^2
= -\det(\X_{ij})$ can be interpreted as measuring the latter (cf. \cite{WC
76}).

\subsection{$N=3$\label{3.3}}

The discussion of the three-dimensional case proceeds along the same lines
as that of the two-dimensional one. We will, therefore, restrict ourselves
to enumerating the relevant points.

\subsubsection*{Constraints}

A basis of the Lie algebra $\LH$ of the gauge group $H=\big( \R^+ \times
SO(3) \big) \semi \R^3$ can be taken to consist of the elements $(D,0)$,
$(K_i,0)$ and $(0,\e_i)$, $i=1,2,3$, where $D=E_3$ and $\{ K_i | (K_i)_{jk}
= -\varepsilon_{ijk}\}$ is the standard basis of $so(3)$. The corresponding
generators of the \c\ algebra ${\cal C}_0$ are
\be \D=x_{ij} p_{ji} + x_i p_i ,\qquad \K_i=\varepsilon_{ijk}\big(-x_{kl}p_{lj}
    + x_j p_k \big) ,\qquad \P_i =p_i. \ee

\subsubsection*{Observables}

For the $sl(3,\R)$ subalgebra of the Lie algebra $cm(3)=sl(3,\R) \semi S(3)$
it is convenient to choose the redundant basis $(T_{ij},0)$, where
\be T_{ij} = E_{ij} - \frac13 \delta_{ij} E_3 ,\qquad\qquad \sum_{i=1}^3
    T_{ii} = 0. \ee
The basis for $S(3)$ can be chosen as above: $S_{ij}=S_{ji}=\frac12 (E_{ij}
+E_{ji})$. The corresponding generators of the algebra \OT\ of fundamental
\obs s are
\be \J_{ij}=\J^R_{(T_{ij},0)} = -x_{kl} (T_{ij})_{lm} p_{mk} ,\qquad\quad
    \X_{ij}=\frac{\lambda}{(\det g)^{2/3}} x_{ki} x_{kj}, \ee
their commutation relations read
\bea \big\{ \J_{ij},\J_{kl} \big\} &=& \delta_{jk} \J_{il} -
                                       \delta_{il} \J_{kj}\\
     \big\{ \J_{ij},\X_{kl} \big\} &=& \delta_{jk} \X_{il} +
                                       \delta_{jl} \X_{ik} -
                                       \frac23 \delta_{ij} \X_{kl}\\
     \big\{ \X_{ij},\X_{kl} \big\} &=& 0.
\eea

\subsubsection*{Casimirs of \O}

The algebra $cm(3)$ possesses two polynomial Casimir invariants (cf.
\cite{WC 76})
\be \CD := \det (\X_{ij}) ,\qquad\quad \CF := -\frac12 \V_k \bar{\V}_k
    \label{cas3}\ee
where
\be \V_k = \varepsilon_{kij} \bar{\X}_{li} \J_{lj} ,\qquad
    \bar{\V}_k = \varepsilon_{kij} \X_{il} \J_{jl} ,\qquad
    \bar{\X}_{ij} = \varepsilon_{ikl} \varepsilon_{jmn} \X_{km} \X_{ln} =
    \bar{\X}_{ji} .\label{vk}\ee
A third invariant is the signature Sig$(\X)$ of the matrix $\X=(\X_{ij})$,
which is defined as twice the number of positive eigenvalues minus the
rank of $\X$.

\subsubsection*{Identities and the \occ}

There are two identities for the Casimirs of \O, which do not involve the
\c s
\be \CD = \det \X \equiv \lambda^3 ,\qquad\qquad {\rm Sig}(\X) =3
    \label{cid3}\ee
(\ie\ $\X$ is positive definite), and one functional identity which
determines the observable content of the \c s
\be \CD \l = \lambda^3 \l \equiv  \CF ,\qquad\qquad
    \l = \sum_{i=1}^3 \bar{\K}_i^2  \label{crel3} \ee
where
\be \bar{\K}_i = \K_i - \varepsilon_{ijk} x_j \P_k . \label{45} \ee
Again, $\l$ is not a Casimir of ${\cal C}_0$ but a generalized \casel\ of
the \c s.
Going over to the equivalent \c\ algebra $\bar{\cal C}_0 = \big( \R^+ \times
SO(3) \big) \times \R^3$ generated by  $\bar{\D} = \D - x_i \P_i$, $\bar{\K}_i$
and $\bar{\P}_i = \P_i$, $\l$ becomes a Casimir of $\bar{\cal C}_0$.

Note that, under the additional conditions expressed by the identities
(\ref{cid3}), the Casimir $\CF$ is always a non-negative function. This
can be seen as follows. In every Hamiltonian action (cf.\ \cite{Wo 80})
of the Lie algebra $cm(3)$ on a
symplectic manifold $M$ the generators of this action are uniquely defined
-- not only up to a constant -- because the first and second cohomology groups
of $cm(3)$ vanish \cite{RI 79}. Therefore, the matrix $\X$ of the generators
$\X_{ij}$ of the $S(3)$ subalgebra assumes the value $\X = \lambda E_3$
(\ie\ there is a point $m \in M$, such that $\X(m) = \lambda E_3$),
given that the generators fulfill the identities (\ref{cid3}). For $\X =
\lambda E_3$ we have $\bar{\X} = 2 \lambda^2 E_3$, and, at the point $m$,
$\CF$ can be written as
\be \CF(m) = \lambda^3 \sum_{i=1}^3 {\sf N}_i^2 (m) ,\qquad\quad
    {\sf N}_i = \varepsilon_{ijk} \J_{jk} .\ee
But $\CF$ is constant on the orbit through $m$, and so $\CF$ is non-negative
everywhere because $M$ is foliated by the orbits of the $cm(3)$ action.

In the case at hand $\CF$ assumes the value zero, the minimum value
which is compatible with the identities (\ref{cid3}), on the constraint
surface. That is, the effect, on the observable sector of the system, of
the vanishing of the constraints consists in restricting the observable
$\lambda^3 \l = \CF$ to its minimum. Consequently, the observable content of
the constraints, \ie, the condition which is imposed on the Casimir $\CF$
of the algebra of observables by the vanishing of the constraints via
the identity (\ref{crel3}), can be given two equivalent formulations,
a numerical and an algebraic one. The numerical formulation states that
the Casimir $\lambda^3 \l = \CF$ assumes the {\it value} zero, whereas the
algebraic formulation states that it has to assume the {\it minimum} possible
value compatible with the identities (\ref{cid3}).

\subsubsection*{Dynamics}

In the same way as in the two-dimensional case the Hamiltonian for the
free constrained dynamics can be expressed as a function of the $cm(3)$
generators. For $q^0=q E_3$ (which is not a restriction because it can
always be achieved by an $SL(3,\R)$--transformation, \ie\ by a change
of basis) it is proportional to the quadratic Casimir of the $sl(3,\R)$
subalgebra
\be \H = \frac1{2q} g_{ijkl} \J_{ji} \J_{lk} ,\qquad\quad g_{ijkl} = {\rm tr}
    \big( T_{ij} T_{kl} \big) .\ee

\section{Quantization of the PRB}

In this section we will carry out the quantization of the PRB in the two-
and three-dimensional case, following the quantization scheme outlined in
Sec.\ II. In both cases we end up with a unique identification of
the physical \r\ of the quantum \aoo.

\subsection{$N=2$}

The first step of the quantization consists in the construction of the Lie
algebra of fundamental observables \QOT\ which corresponds to the classical
algebra \OT\ on the level of quantum theory. The only quantum correction
of the classical commutation relations, compatible with the principles
formulated in Sec.\ II, would be a central extension of the algebra
$iso(2,1)$. However, the Lie algebra $iso(2,1)$ does not possess any
non-trivial central extension, because its second cohomology vanishes
(cf.\ \cite{Wo 80}). Thus, the Lie algebraic structure of the algebra of
fundamental \obs s remains unchanged, \ie\ the algebra \QOT, as a
commutator algebra, is isomorphic to the Lie algebra $iso(2,1)$, and
the algebra of \obs s \QO\ is isomorphic to its enveloping algebra.
The commutation relations of the generators $\hat{\M}_{\mu\nu}$ and
$\hat{\X}_\mu$ of \QOT\ read
\be \big[ \hat{\M}_{\mu\nu},\hat{\M}_{\mu\rho} \big] = i\hbar g_{\mu\mu}
    \hat{\M}_{\nu\rho} ,\qquad
    \big[ \hat{\M}_{\mu\nu},\hat{\X}_\mu \big] = i\hbar g_{\mu\mu}
    \hat{\X}_\nu ,\qquad
    \big[ \hat{\X}_\mu,\hat{\X}_\nu \big] = 0. \ee
The expressions for the Casimirs of \QO\ in terms of $\hat{\M}_{\mu\nu}$ and
$\hat{\X}_\mu$ are the same as in eqs.\ (\ref{cas2}). Observe that there
are no factor ordering ambiguities in the definition of the Casimirs. Of
course, the expressions for other \obs s, in terms of the basic $iso(2,1)$
\obs s, and their commutation relations can still acquire quantum corrections.

Next we have to determine the form which the identities (\ref{cid2}) for the
Casimirs of $iso(2,1)$ take upon quantization. Imposing the requirements
that the classical identities be reproduced in the classical limit, that
the possible correction terms must carry
explicit positive integer powers of $\hbar$ and the correct overall physical
dimensions, and that the identities may only involve Casimirs, it can easily
be seen that there are no quantum corrections available. That is, the form of
the identities remains unchanged
\be \hat{\X}^2 \equiv -\lambda^2 ,\qquad\qquad {\rm sign}(\hat{\X}_3) = 1
    \label{qid2}\ee
(the latter identity means that $\hat{\X}_3$ is a positive operator).

The crucial step is the identification of that Casimir element in the algebra
of \obs s, which corresponds to the classical \obs\ $\bar{\K}$. Applying the
same principles as above, it can be seen that the only possible correction of
the classical expression is a constant term
\be \lambda \hat{\bar{\K}} = \hat{{\sf I}}_2 + \hbar\lambda c ,\qquad\quad
    c = \mbox{const}. \label{corr2}\ee
This constant contribution can be excluded by an additional discrete
symmetry (parity) of the classical system. Implementing it as a reflection
symmetry, the cooordinates transform as
\be (x_i,p_i) \longrightarrow (-1)^i (x_i,p_i), \quad\qquad
    (x_{ij},p_{ij}) \longrightarrow (-1)^{i+j} (x_{ij},p_{ij}). \ee
The classical \obs s $\CZ$ and $\bar{\K}$ transform as pseudoscalars.
If we require this symmetry to be realized in the quantum theory, the
constant $c$ must vanish. Therefore, we have the identification
\be \lambda \hat{\bar{\K}} = \hat{{\sf I}}_2 \label{qrel2},\ee
and the \occ\ is expressed by the induced operator identity $\hat{\sf I}_2=0$.

\subsubsection*{Identification of the physical \r\ of \QO}

As the Hermitian irreducible \r s (HIR) of the Lie algebra $iso(2,1)$ can
be obtained from the unitary irreducible \r s (UIR) of the group $ISO(2,1)$,
we will in the sequel determine the physical \r\ of the latter. This can
most easily be done by using the method of induced \r s (cf. \cite{BR 80}).

First, the identities (\ref{qid2}) fix an orbit under the action of
$SO(2,1)$ on the dual space of the space of characters of the Abelian
subgroup $\R^3$. This orbit can be identified with the ``mass" shell
\be H_{2,1}^+(\lambda) = \Big\{ \x \in \R^3 \Big| \x^2 = g^{\mu\nu}
    x_\mu x_\nu = -\lambda^2 ,\; x_3 > 0 \Big\} \ee
and is isomorphic to the homogeneous space $SO(2)\big\backslash SO(2,1)$.
This means that the
identities (\ref{qid2}) single out a class of \r s of $ISO(2,1)$, namely
those \r s  which are induced by the UIR of the subgroup $SO(2) \semi \R^3$
associated with the orbit $H_{2,1}^+(\lambda)$. The individual \r s in this
class are characterized by the corresponding \r\ of the little group
$SO(2)$, labelled by an integer $k_0 \in {\bf Z}$. They will be denoted
D$_2(\lambda,k_0)$.

The eigenvalue of the Casimir $\hat{{\sf I}}_2$ in the \r\ D$_2(\lambda,
k_0)$ is given by $\hat{{\sf I}}_2 = \hbar \lambda k_0$. Consequently, the
identity $\hat{\sf I}_2 = \lambda \hat{\bar{\K}} = 0$, acting as a \r\
condition, uniquely selects the \r\ D$_2(\lambda,0)$ as the physical \r\
of the group $ISO(2,1)$ and of the algebra \QOT\ of fundamental \obs s.

Finally, we want to give an explicit description of the physical \r\ of the
group $ISO(2,1)$, respectively of the equivalent \r\ of $CM(2)$, and to
determine the spectrum of the free Hamiltonian.

The UIR of $ISO(2,1)$ which is induced by the spin-zero \r\ of the subgroup
$SO(2) \semi \R^3$ associated with the orbit $H^+_{2,1}(\lambda)$ is realized
on the Hilbert space
\be {\cal H}_{2,1}=L^2 \big( H^+_{2,1}(\lambda),d\mu(x) \big) ,\qquad\qquad
    d\mu(x) = \delta(\x^2 + \lambda^2) \theta (x_3) d^3x. \ee
An element $(g,\a) \in ISO(2,1)$ acts on ${\cal H}_{2,1}$ via the unitary
operator $\hat{\U}(g,\a)$
\be \big( \hat{\U}(g,\a) \Psi \big) (\x) = e^{\frac{i}\hbar \a \cdot \x}
    \Psi(g^{-1}\x) \ee
($\a \cdot \x = x_\mu a^\mu$). This \r\ is unitarily equivalent to the
following UIR of $CM(2)$, induced by the spin-zero \r\ of the subgroup
$SO(2) \semi S(2)$ related to the orbit
\[O_2(\lambda) = \Big\{ X\in S(2) \Big| \det X = \lambda^2, \: \mbox{$X$
  positive definite} \Big\} ,\]
realized on the Hilbert space
\[ {\cal H}_2 = L^2 \big( O_2(\lambda),d\mu(X) \big) \]
where
\be d\mu(X) = \delta(\det X - \lambda^2) \theta(X) dX_{11} dX_{12} dX_{22} \ee
\be \theta(X)= \left\{ \begin{array}{cl} 1 & \mbox{if $X$ is positive definite}
    \\ 0 & \mbox{else.} \end{array} \right. \ee
The action of an element $(g,S) \in CM(2)$ on ${\cal H}_2$ is given by
\be \big( \hat{\U}(g,S) \Psi \big) (X) = e^{\frac{i}\hbar \, {\rm tr}(SX)}
    \Psi(g^t X g). \ee
Upon restriction to the subgroup $SO(2,1)$ the above \r\ of $ISO(2,1)$
decomposes into a direct integral
\[ \int_0^\infty D(\sigma) d\mu(\sigma) \]
of the \r s $D(\sigma)$ of the continuous series of UIR of $SO(2,1)$ (cf.\
\cite{VK 93}). The parameter $\sigma$ which characterizes the \r s is
connected to the eigenvalues of the quadratic Casimir $\hat{\L}^2$ of
$so(2,1)$ in these \r s via
\be \mbox{spec} \big( \hat{\L}^2 \big) = \Big\{ \hbar^2 \big( \frac14 +
    \sigma^2 \big) \Big| \sigma \in \R^+ \Big\} .\ee
Therefore, assuming that the classical relation (\ref{ham2}) between the
free Hamiltonian and $\L^2$ remains unchanged because there are no factor
ordering problems for $\hat{\H}$ in terms of the fundamental \obs s
$\hat{\L}_\mu$ (a constant quantum correction $\hbar^2 \frac{c}{2 \theta}$
is compatible with the correspondence principle, but it cannot be measured),
the spectrum of $\hat{\H}$ is given by
\be \mbox{spec} \big( \hat{\H} \big) = \Big\{ \frac1{2\theta} \hbar^2
    \big( \frac14 + \sigma^2 \big) \Big| \sigma > 0 \Big\} .\ee

\subsection{$N=3$\label{4.2}}

Applying the same principles and argumentation as in the two-dimensional
case, the Lie algebra \QOT\ has to be taken to be isomorphic to the Lie
algebra $cm(3)$. The commutation relations of the generators $\hat{\J}_{ij}$
and $\hat{\X}_{kl}$ read
\bea \big[ \hat{\J}_{ij},\hat{\J}_{kl} \big] &=& i\hbar \big( \delta_{jk}
     \hat{\J}_{il} - \delta_{il} \hat{\J}_{kj} \big)\\
     \big[ \hat{\J}_{ij},\hat{\X}_{kl} \big] &=& i\hbar \big( \delta_{jk}
     \hat{\X}_{il} + \delta_{jl} \hat{\X}_{ik} -
     \frac23 \delta_{ij} \hat{\X}_{kl} \big)\\
     \big[ \hat{\X}_{ij},\hat{\X}_{kl} \big] &=& 0.
\eea
The expressions for the Casimirs of $cm(3)$ in terms of the generators
$\hat{\J}_{ij}$ and $\hat{\X}_{ij}$ are the same as the classical ones
(eq.\ (\ref{cas3})). Note that, because of the identities
\[ \hat{\V}_k=\varepsilon_{kij} \hat{\bar{\X}}_{li} \hat{\J}_{lj}
             =\varepsilon_{kij} \hat{\J}_{lj} \hat{\bar{\X}}_{li} ,\qquad
   \hat{\bar{\V}}_k=\varepsilon_{kij} \hat{\X}_{il} \hat{\J}_{jl}
                   =\varepsilon_{kij} \hat{\J}_{jl} \hat{\X}_{il} ,\qquad
   \hat{\V}_k \hat{\bar{\V}}_k = \hat{\bar{\V}}_k \hat{\V}_k, \]
the Casimir $\hat{{\sf I}}_5 = - \frac12 \hat{\V}_k \hat{\bar{\V}}_k$ is
unambiguously defined by its classical expression. The form of the identities
(\ref{cid3}) remains unchanged
\be \hat{{\sf I}}_3 = \det \hat{\X} \equiv \lambda^3 , \qquad\qquad
    {\rm Sig}(\hat{\X}) = 3  \label{qid3}\ee
and the identification of the Casimir element expressing the \obs\
content of the \c s can be accomplished up to a contribution proportional
to $\hat{{\sf I}}_3$
\be \hat{{\sf I}}_3 \hat{\l} = \lambda^3 \hat{\l} =
    \hat{{\sf I}}_5 + c \hbar^2 \hat{{\sf I}}_3 = \hat{{\sf I}}_5 +
    c \hbar^2 \lambda^3 ,\qquad\quad c = \mbox{const.} \label{corr3}\ee
The identities (\ref{qid3}) determine a class of UIR of the group $CM(3)$.
These \r s are induced by the UIR of the subgroup $SO(3)\semi S(3)$ 
associated with the orbit
\[ O_3(\lambda) = \Big\{X \in S(3) \Big| \det X = \lambda^3, \: X \mbox{
   positive definite} \Big\}. \]
$O_3(\lambda)$ is isomorphic to the homogeneous space $(SO(3) \semi S(3))
\big\backslash CM(3)$. The \r s are labelled by a discrete parameter $j$,
$2j \in {\bf N}_0$, which characterizes the corresponding \r\ of the
little group $SO(3)$ (respectively of its covering group $SU(2)$). They will
be denoted D$_3 (\lambda,j)$. The eigenvalues of the Casimir $\hat{{\sf I}}_5$
in these \r s are given by (cf.\ \cite{WC 76})
\be \hat{{\sf I}}_5 = \hbar^2 \lambda^3 j(j+1) .\ee
    
Applying the algebraic formulation of the observable content of the
constraints given in Sec.\ IV.\ref{3.3}, the implementation of the \c s
can be carried out without explicitly having to determine the constant $c$
in eq.\ (\ref{corr3}). Accordingly, the physical representation is
distinguished by the fact that the observable $\hat{\l}$ assumes the minimum
value compatible with the identities (\ref{qid3}). As in the classical case,
under the additional conditions expressed by (\ref{qid3}), the Casimir
$\hat{{\sf I}}_5$ is non-negative. It assumes its minimum value zero in the
representation D$_3(\lambda,0)$. Therefore, irrespective of the value
of the constant $c$, the physical representation of $CM(3)$ can be
identified uniquely as the representation D$_3(\lambda,0)$. For $c=0$,
$\hat{\l}$ assumes the value zero in this representation.

The physical \r\ is realized on the Hilbert space
\be {\cal H}_3 = L^2 \big( O_3(\lambda), \: d\mu(X) \big) , \quad
    d\mu(X)= \delta(\det X - \lambda^3) \theta(X) \prod_{i \le j} dX_{ij} \ee
with the group action being given by $(g,S) \longmapsto \hat{\U}(g,S)$
\be \big( \hat{\U}(g,S) \Psi \big) (X) = e^{\frac{i}\hbar \,{\rm tr}(SX)}
    \Psi(g^t X g).\ee
As a \r\ space for $CM(3)$, the Hilbert space ${\cal H}_3$ coincides with the
one that has been determined in Ref. \cite{DE 90}.

Upon restriction to $SL(3,\R)$ the \r\ D$_3(\lambda,0)$ decomposes into a
direct integral of UIR of $SL(3,\R)$. This direct integral decomposition
yields the spectral resolution of the Hamiltonian $\hat{\H}$, which (again
up to an overall additive constant $\hbar^2 \frac{c}{2q}$) can be identified
as being proportional to the second order \cas\ of $sl(3,\R)$
\[ \hat\H = \frac1{2q} g_{ijkl} \hat\J_{ji} \hat\J_{lk} \]
(because of the identity $g_{ijkl} \hat\J_{ji} \hat\J_{lk} = g_{ijkl}
\hat\J_{lk} \hat\J_{ji}$ there are no factor ordering ambiguities).
Unfortunately, the required decomposition and the corresponding eigenvalues
of the \casop s are not available in the standard mathematical literature on
the subject, so that we cannot determine the spectrum of the Hamiltonian
explicitly.

\newpage
\section{Conclusions}

In this paper I have demonstrated the usefulness and effectiveness of our
algebraic \c\ quantization scheme for the construction of the quantum theory
of a system with a complicated, non-Abelian and non-unimodular, gauge group.
By the discussion of this example the algebraic method could be given a
more precise and more widely applicable formulation.

I would like to stress the importance of the proper identification of the
\obs\ content of the \c s. It allows to carry over the implementation of the
\c s to the \obs\ sector and to deal only with \obs\ quantities. Thus it
can be avoided to make the quantization of a constrained system depend on
manipulations in the un\obs\ sector of the \c s. Once the \obs\ content of
the \c s has been determined classically, the \c s are discarded and the
construction of the quantum theory proceeds by applying correspondence and
consistency requirements to observable quantities. And, after all, these
principles can only be applied to observable quantities. There are no
guiding principles for the construction of a quantum theory of un\obs\
quantities.

In the case of the pseudo-rigid body this feature of the algebraic \c\
quantization scheme leads to a considerable simplification, in that it is
not necessary to take care of the non-unimodularity of the gauge group,
caused by the presence of the dilation \c. In the conventional approaches
to the quantization of constrained systems this structure of the gauge
group leads to serious difficulties.

Future applications of our algebraic \c\ quantization scheme will be
concerned with the proper implementation of the \c s -- and not of gauge
conditions -- of physically relevant relativistic field theories into
the respective quantum theories.


\begin{thebibliography}{99}
\bibitem{Anm 1}  In the sequel the term ``\c s" will be used for the
                 generators of infinitesimal gauge transformations as well
                 as for the constraining conditions themselves. Its precise
                 meaning will be clear from the context. We will only
                 consider systems with {\it first class} \c s \cite{Di 64}.
\bibitem{Anm 3}  Where the \casinv s, or simply Casimirs, comprise the proper
                 \casop s, \ie\ the generators of the center of the \aoo, as
                 well as all further invariant quantities which can be used
                 for the characterization of (classes of) \ir s, like, \eg,
                 the sign of the energy in the physical \r s of the Poincar\'e
                 algebra.
\bibitem{Anm 4}  As in the case of the \c s, this term will be used not only
                 for the conditions themselves, but also for the gauge
                 invariant combinations of the \c s, respectively the
                 corresponding \casel s of the \aoo, which enter into
                 the functional dependencies.
\bibitem{Anm 5}  In some cases it may be more favourable that not all the
                 generators of \S\ are included in \OT\ but, \eg, only
                 the generators of the space--time symmetries. In that case
                 those elements of \S\ which are not contained in \OT\
                 must be simple polynomial functions in the elements of
                 \OT\ and \S\ must be treated separately in the same way
                 as \OT\ respectively \O\ (\cf\ Ref.\ \cite{Tr 96}).
\end{thebibliography}
\end{document}